\documentclass[%
reprint,
superscriptaddress,
amsmath,amssymb,
aps,
prb,
floatfix,
]{revtex4-2}

\usepackage{multirow}
\usepackage{graphicx}% Include figure files
\usepackage{dcolumn}% Align table columns on decimal point
\usepackage{bm}% bold math
\usepackage[version=3]{mhchem}
\usepackage{nicefrac}

\usepackage{color}
\usepackage{xcolor}
\usepackage{dcolumn}
\usepackage{amssymb}
\usepackage{url}
\usepackage{hyperref}
\usepackage{xfrac}
\usepackage{tabularx}
\usepackage{xspace}
\usepackage{amsmath}
\usepackage[normalem]{ulem}
\usepackage{mathtools}
\usepackage{physics}
\usepackage{dsfont}
\usepackage{float}
\usepackage{braket}
\usepackage{placeins}
\usepackage{footmisc}

\def\be{\begin{equation}}
\def\ee{\end{equation}}
\def\bea{\begin{eqnarray}}
\def\eea{\end{eqnarray}}
\def\ba{\begin{array}}
\def\ea{\end{array}}

\begin{document}

%\title{Competing valence bond solids in the ground state of the spin-$1/2$ Heisenberg antiferromagnet on the star lattice}
%\title{Competing valence bond-solid phases in the spin-1/2 Heisenberg antiferromagnet on the star lattice: a tensor network and series expansion study}
\title{Tensor-network study of the ground state of maple-leaf Heisenberg antiferromagnet}

\author{Samuel Nyckees}
%\thanks{These authors contributed equally.}
\affiliation{Institute of Physics, Ecole Polytechnique Fédérale de Lausanne (EPFL), CH-1015 Lausanne, Switzerland}
\affiliation{Now at Universit\'e Grenoble Alpes, CEA List, 38000 Grenoble, France}

\author{Pratyay Ghosh}
%\thanks{These authors contributed equally.}
\email{pratyay.ghosh@epfl.ch}
\affiliation{Institute of Physics, Ecole Polytechnique Fédérale de Lausanne (EPFL), CH-1015 Lausanne, Switzerland}

\author{Fr\'ed\'eric Mila}
%\thanks{These authors contributed equally.}
\affiliation{Institute of Physics, Ecole Polytechnique Fédérale de Lausanne (EPFL), CH-1015 Lausanne, Switzerland}

\begin{abstract}
We study the quantum phase diagram of the spin-$1/2$ nearest-neighbor Heisenberg model on the maple-leaf lattice using infinite projected entangled pair states (iPEPS) combined with a corner transfer matrix renormalization group scheme adapted to $C_3$-symmetric lattices. Focusing on the fully antiferromagnetic $J$-$J_d$ model with $J_h = J_t := J$, we map out the ground-state phase diagram as a function of the dimer coupling $J_d$. Our results show that the system hosts only two phases: a magnetically ordered canted-$120^\circ$ phase and an exact dimer singlet product phase. We identify a first-order transition between these two phases at $J_d/J \approx 1.45$. Within the magnetically ordered phase, we observe small but finite magnetic moments. We also resolve the quantum renormalization of the canting angle, which deviates from the classical prediction over almost the entire magnetically ordered phase.
\end{abstract}

\maketitle

\section{Introduction}
Quantum antiferromagnets on frustrated lattices are prototypical systems in which competing interactions or lattice geometry prevent spins from simultaneously satisfying all constraints, often resulting in highly degenerate ground states and unconventional magnetic or exotic quantum-disordered phases~\cite{frustrationbook,Diepbook}.
Against this backdrop, the $S=1/2$ Heisenberg antiferromagnet on the maple-leaf lattice (MLL)~\cite{Betts1995} has recently become an active topic of investigations due to its ability to host an exact dimer product ground state~\cite{Farnell2011,Ghosh2022}, spin liquid phases~\cite{Gresista2023,Ghosh2024-al,schmoll2024bathingseacandidatequantum}, unusual canted magnetic order~\cite{Farnell2011,Ghosh2022,Gembe_mll}, and predicted exotic magnetization plateaus formed by multi-particle bound states~\cite{Ghosh2023}. 
Moreover, several material realizations exist in which the underlying lattice of magnetic ions follow the maple-leaf structure~\cite{Haraguchi2021,Lake2025,Fennel2011,Schmoll2025,Ghosh2023b,Ghosh2025}, further enhancing interest in the magnetic properties of this system. 

\begin{figure}[t]
    \includegraphics[width=0.95\columnwidth]{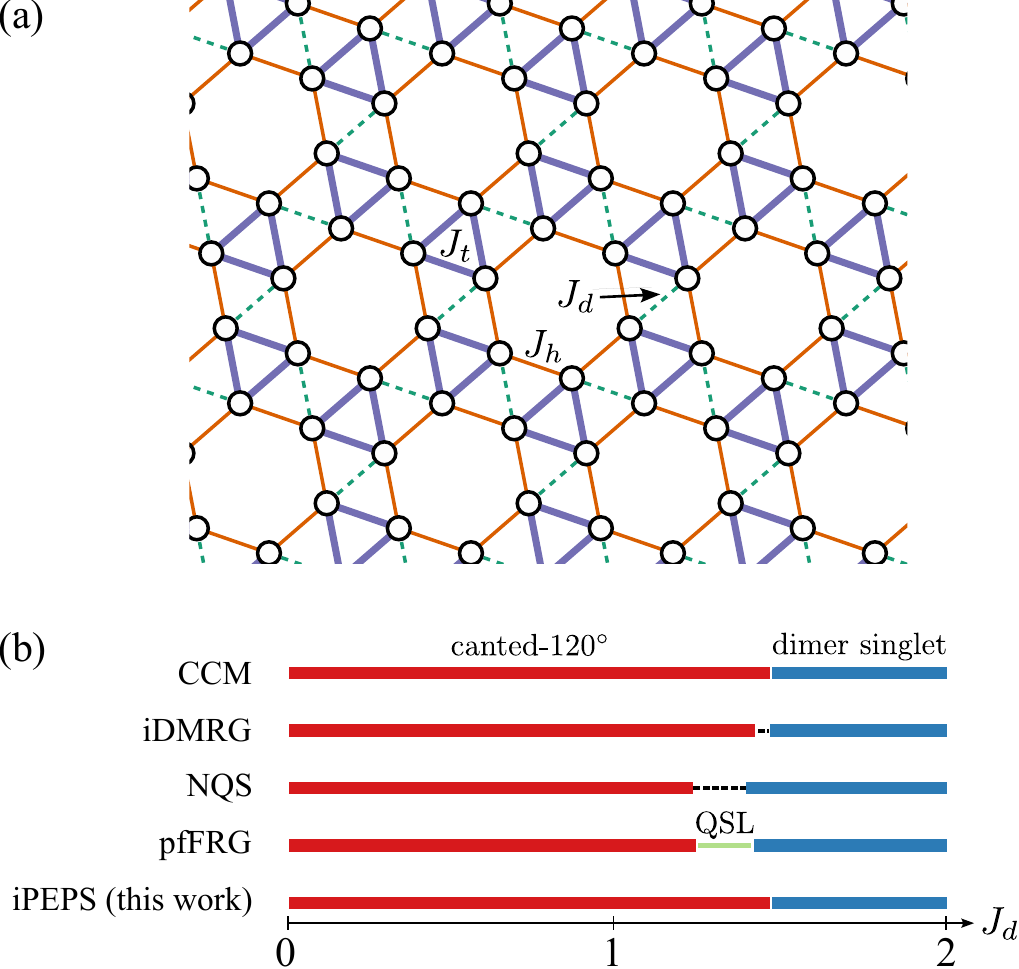}
    \caption{(a) Maple-leaf lattice (MLL) with nearest-neighbor interactions. The green dashed bonds denote the dimer couplings $J_d$, while the thick violet and thin red bonds represent the inter-dimer couplings $J_t$ and $J_h$, respectively. (b) Ground-state phase diagram of the Heisenberg model on the maple-leaf lattice with $J_h = J_t = 1$ as a function of $J_d$, obtained using different numerical techniques, including the present work. Both CCM~\cite{Farnell2011} and our iPEPS simulations find that the system realizes only two phases, namely the canted-$120^\circ$ ordered phase and the product dimer singlet phase, separated by a first-order transition at $J_d \approx 1.45$. The iDMRG~\cite{Beck2024} and NQS calculations~\cite{Beck2024} also identify these two phases, but with an intervening region of uncertainty between them, indicated by the dashed lines. In contrast, the pseudofermion functional renormalization group study~\cite{Gresista2023} reports a quantum spin liquid phase between the canted-$120^\circ$ ordered phase and the product dimer singlet phase.} \label{fig-lattice}
\end{figure}

\begin{figure*}[t]
    \includegraphics[width=0.98\textwidth]{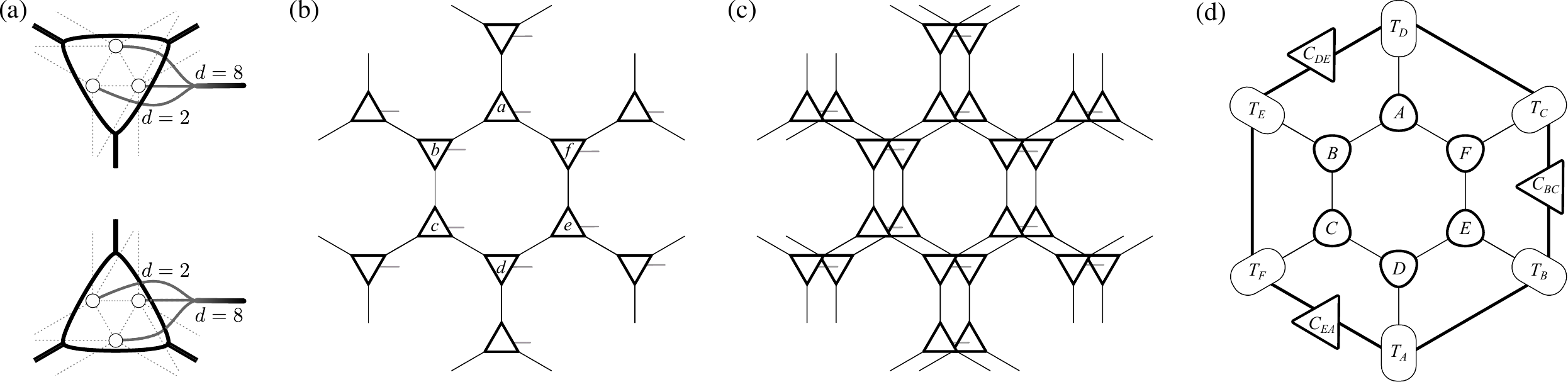}
    \caption{(a) Coarse-grained three-site tensors. Each $J_t$ trimer of the lattice is represented by a tensor with physical dimension $d = 8$. The thick black lines indicate virtual bonds of dimension $D$.  
    (b) Two-dimensional iPEPS wave function defined on a honeycomb network with a six-site unit cell. The gray lines denote the physical legs.  
    (c) The wave-function overlap is represented by the contraction of an infinite tensor network.  
    (d) This contraction is approximated by an effective environment composed of row and corner tensors. The thicker black lines indicate environment bonds of dimension $\chi$.
} \label{fig-TN}
\end{figure*}

The MLL [see Fig.~\ref{fig-lattice}(a)] is one of the eleven Archimedean lattices and has a coordination number of five. It can be viewed as one-seventh site-depleted (one-fifth bond-depleted) triangular lattice~\cite{Schulenburg2000,maple_richter,Farnell2011}. The lattice contains three symmetry-inequivalent nearest-neighbor bonds, which can be assigned three different interaction strengths: $J_h$ on the hexagons, $J_t$ on the triangles, and $J_d$ on the dimers, as shown in Fig.~\ref{fig-lattice}(a). The spin-$1/2$ Heisenberg model on the MLL is thus given by
\be\label{eq:hmail}
H=\sum_{\langle ij \rangle_k} J_k \mathbf{S}_i \cdot \mathbf{S}_j ,
\ee
where $\mathbf{S}_i$ are spin-$1/2$ operators and $\langle ij \rangle_k$ denotes a nearest-neighbor bond of type $k \in \{h,t,d\}$. A central focus of theoretical investigations of the fully antiferromagnetic model on the MLL is the case $J_h = J_t := J$, with the phase diagram explored as a function of $J_d$. The earliest study of this model was carried out by Richter \emph{et al.} using the coupled cluster method (CCM) and exact diagonalization~\cite{Farnell2011}. Their results suggest that the phase diagram hosts only two phases, namely an unusual magnetically ordered phase, identified as canted-$120^\circ$ order, and an exact dimer singlet product phase. The phase transition between these two phases was found to occur at $J_d/J \approx 1.45$~\cite{Farnell2011}. Subsequent studies employing pseudofermion functional renormalization group (pfFRG)~\cite{Gresista2023}, infinite-density matrix renormalization group (iDMRG)~\cite{Beck2024}, and neural quantum states (NQS)~\cite{Beck2024} are largely consistent with these findings. 
These studies, however, also raise the possibility of an intermediate phase located between the canted-$120^\circ$ ordered phase and the exact dimer phase; the iDMRG identify an intervening valence bond-like phase whose true nature remains unresolved, whereas the pfFRG approach reports a quantum spin liquid (QSL) ground phase. A comprehensive overview of the phase diagrams obtained from different numerical techniques (together with the results of the present work) is shown in Fig.~\ref{fig-lattice}(b).

%Another important direction in the study of this model involves the exact singlet state under an applied Zeeman field ($h$), which is expected to produce a sequence of superfluid to insulator transitions that appear as magnetization plateaus~\cite{Ghosh2023}. In Ref.~\cite{Farnell2011}, this problem was investigated this using ED, which remains the only work to date that explores the full range of magnetic fields in this model. The authors identified two magnetization plateaus at $m/m_s = 1/3$ and $2/3$, although these are finite size results.

In this article, we examine the spin-$1/2$ nearest-neighbor antiferromagnetic Heisenberg model on the MLL. We set $J = 1$ throughout this work and investigate the system as a function of $J_d$ using the state-of-the-art tensor network method of infinite projected entangled pair states (iPEPS), together with a specialized corner transfer matrix renormalization group scheme adapted to $C_3$-symmetric systems to approximate the environment~\cite{gendiar2012,nyckees2023,lukin2023,lukin2024,star_ghosh,ghosh_ruby}. iPEPS is a variational ansatz in which the wave function of a two-dimensional infinite system is represented through a network of tensors~\cite{Jiang2008,Jordan2008,Verstraete2004a,Verstraete2004b,Gu2008}, allowing direct access to the ground state properties in the thermodynamic limit. Our iPEPS simulations show that the ground-state phase diagram contains only the canted-$120^\circ$ ordered phase and the exact dimer singlet phase, in agreement with the findings of CCM~\cite{Farnell2011}, with no evidence for an intermediate phase. We find a first-order phase transition between these two phases at $J_d \approx 1.45$. 
%In the presence of a magnetic field, we find the system enters $m/m_s = 1/3$ and $2/3$ plateaus confirming the expectation from the ED study. Moreover, we find the evidence of additional plateaus in the system, namely at $m/m_s = 1/9$ and $2/9$. 

The remainder of the article is organized as follows. In Sec.~\ref{sec:methods}, we describe the iPEPS methodology used to study the maple-leaf Heisenberg antiferromagnet. In Sec.~\ref{sec:result_gs}, we present the ground-state phase diagram of the model. Finally, Sec.~\ref{sec:conclusions} summarizes our main findings and outlines directions for future research.
%This is followed by the details of the magnetization behavior within the exact dimer phase in Sec.~\ref{sec:result_mag}. 

\begin{figure*}[t]
    \includegraphics[width=0.98\textwidth]{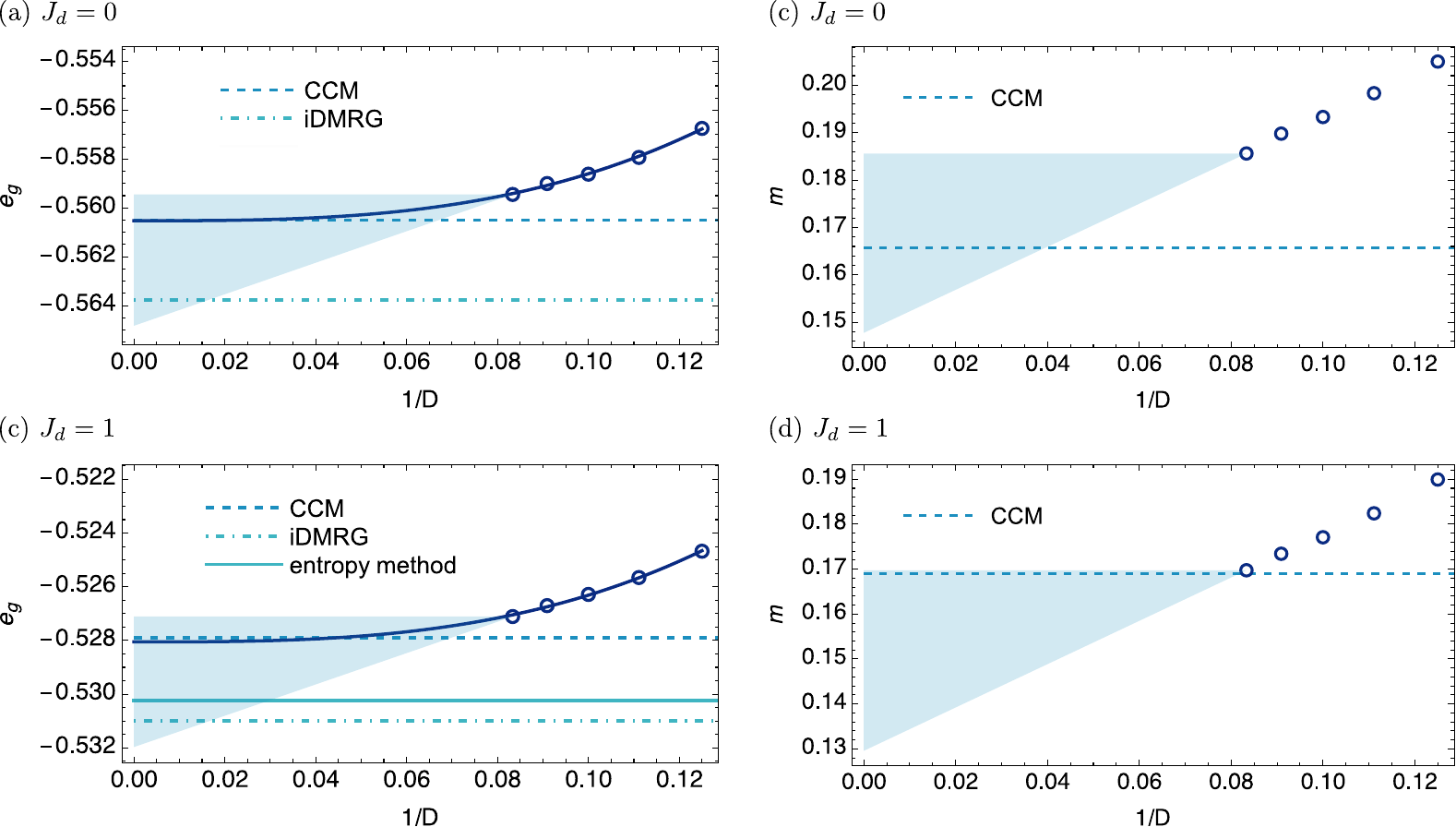}
    \caption{Scaling of the ground-state energy per site, $e_g$, obtained from iPEPS as a function of $1/D$ at $J_h = J_t = 1$ for (a) $J_d = 0$ and (b) $J_d = 1$. Energies obtained using different numerical approaches, namely CCM~\cite{Farnell2011}, iDMRG~\cite{Beck2024}, and entropy method~\cite{Hutsak_maple}, are also indicated. The dark solid lines show the $D \to \infty$ extrapolations obtained by fitting the data. The shaded regions represent estimates of the error bars, where the upper bounds correspond to the energies at $D = 12$, and the lower bounds are obtained from a linear extrapolation of the $D = 11$ and $D = 12$ data. Scaling of the average local magnetic moment, $m$, obtained from iPEPS as a function of $1/D$ at $J_h = J_t = 1$ for (c) $J_d = 0$ and (d) $J_d = 1$. Conservative error bar estimates for $m$, indicated by the shaded regions, are obtained in the same manner as for the energy.  
    %The dark solid lines denote linear extrapolations of the data.
} \label{fig-gs_mag_1}
\end{figure*}

\section{Infinite projected entangled pair states (iPEPS)}\label{sec:methods}
The iPEPS ansatz represents the ground-state wave function of a two-dimensional system as an infinite two-dimensional tensor network. In this work, we consider a tensor network in which each $J_t$ trimer of the MLL is represented by a single tensor with a physical leg of dimension $d = 2^3$ [Fig.~\ref{fig-TN}(a)]. This construction leads to a representation of the ground state as a tensor network defined on the honeycomb lattice, with tensors $a^{[\mathbf{x}]}_{i,j,k,s}$ [see Fig.~\ref{fig-TN}(b)], where the indices $i,j,k$ label the virtual spaces of dimension $D$, and the index $s$ corresponds to the physical space. The position of each tensor within the unit cell is denoted by $[\mathbf{x}]$.

To obtain the ground state, the tensors $a^{[\mathbf{x}]}_{i,j,k,s}$ are optimized by minimizing the energy of the system. Various optimization strategies exists, including the simple update~\cite{Jiang2008}, full update~\cite{Jordan2008,Phien2015}, and automatic differentiation~\cite{Hasik2021}. In this work, we employ the simple update algorithm, in which a random initial tensor network $| \psi_0 \rangle$ is evolved in imaginary time through a sequence of local, quasi-adiabatic updates, mimicking a slow annealing process that relaxes the system toward the ground state $| \psi \rangle$. For gapped systems, where entanglement is relatively short-ranged, this approach provides an efficient and accurate approximation of the ground state. The imaginary-time evolution is implemented by iteratively applying a Trotter–Suzuki decomposition of $e^{-\tau H}$,
\begin{align}
| \psi \rangle & = \lim_{\beta \rightarrow \infty} e^{-\beta H} | \psi_0 \rangle \ \\
& \simeq \lim_{n \rightarrow \infty} \left(\prod_{e \in \langle i, j \rangle} e^{-\tau H_{e}}\right)^n | \psi_0 \rangle  .
\end{align}
In practice, we use $\tau = 10^{-3}$–$10^{-2}$. The imaginary-time evolution is carried out using two-site gates that are successively applied to nearest-neighbor tensors. After each gate application, the local tensors $a^{[\mathbf{x}]}$ are projected back onto a relevant subspace of dimension $D$ by performing a singular value decomposition and retaining only the $D$ largest singular values.

Once the local tensors have been optimized, local observables are evaluated by contracting the infinite two-dimensional tensor network defined in terms of the local tensor
$$A_{i^{},i',j^{},j',k^{},k'}^{[\mathbf{x}]} = a_{i',j',k',s}^{[\mathbf{x}]\dagger} a_{i^{},j^{},k^{},s^{}}^{[\mathbf{x}]}$$
see Fig.~\ref{fig-TN}(c). To perform this contraction, we employ the corner transfer matrix renormalization group (CTMRG) algorithm specifically designed for tensor networks on the honeycomb lattice~\cite{gendiar2012,nyckees2023,lukin2023,lukin2024,star_ghosh,ghosh_ruby}. The CTMRG approximates the infinite contraction using an effective environment composed of row tensors $T^{[\mathbf{x}]}$ and corner tensors $C^{[\mathbf{x}]}$, as illustrated in Fig.~\ref{fig-TN}(d), with dimensions $\chi \times D^2 \times \chi$ and $\chi \times \chi$, respectively. The parameter $\chi$ serves as a control parameter, and one expects the contraction to converge to the exact result in the limit $\chi \rightarrow \infty$. In practice, it is necessary to verify that observables, such as the energy, are converged with respect to $\chi$. In this study, we use $\chi$ up to $120$, for which the energy has sufficiently converged (with a precision of $10^{-6}$) for all bond dimensions $D \in [8,12]$.

\begin{figure*}[t]
    \includegraphics[width=0.98\textwidth]{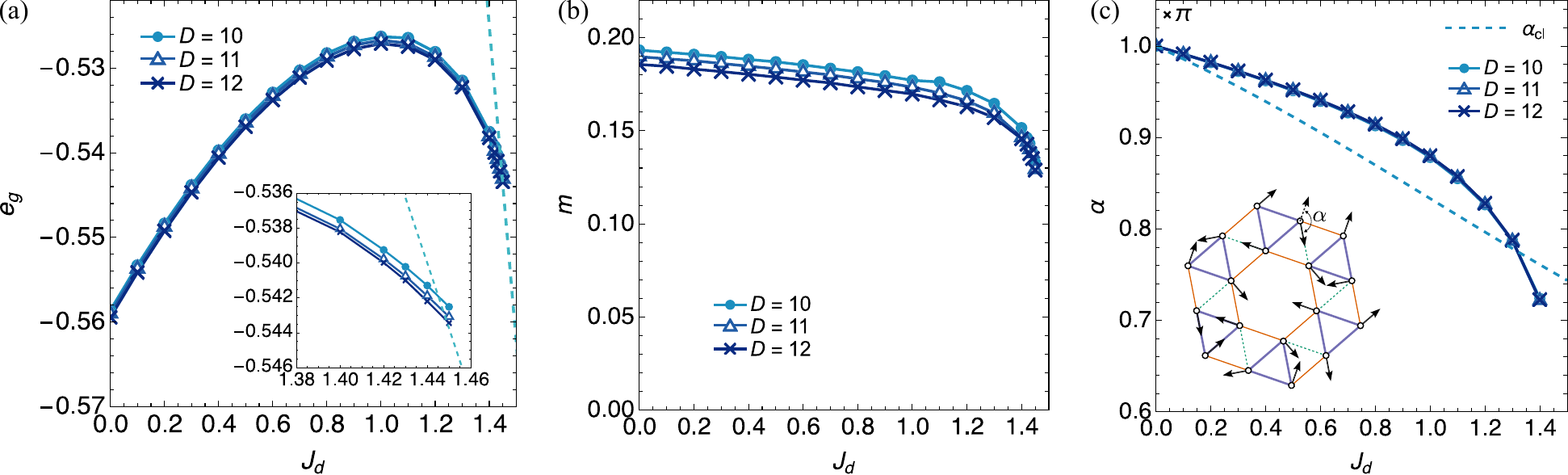}
    \caption{(a) Energy of the canted-$120^\circ$ ordered ground state as a function of $J_d$, obtained from iPEPS calculations for different bond dimensions $D$. The dashed line indicates the energy of the exact dimer singlet product state. The transition between the canted-$120^\circ$ phase and the exact dimer phase is clearly first order, as can be seen in the inset.  
    (b) Dependence of the local spin moment, $m$, in the canted-$120^\circ$ phase on $J_d$, obtained from iPEPS calculations for various bond dimensions $D$.  
    (c) Canting angle $\alpha$ as a function of $J_d$. The definition of the canting angle is shown in the inset. The dashed curve corresponds to the classical canting angle $\alpha_\text{cl}$ [see Eq.~\eqref{eq:alpha}].
} \label{fig-gs_mag_2}
\end{figure*}

%\begin{figure*}[t]
%    \includegraphics[width=0.98\textwidth]{fig_GS.pdf}
%    \caption{(a) Scaling of the ground state energies per site, $E$, obtained in iPEPS as a function of $1/D$ at $J_h=J_t=1$ and $J_d=0$ and $J_d=1$ \changes{@Sam: (1) We should only compare with the extrapolated values from CCM (see Table 1)}. The solid line shows the extrapolation for $D\to\infty$ obtained via an exponential fitting of the data obtained for $D=8$-$12$. \changes{@Sam We should do this extrapolation?} The horizontal dashed lines corresponds to the energies reported in Ref.\cite{Farnell2011} obtained via CCM calculations. (b) Energies as a function of $J_d$ obtained from iPEPS calculations for various bond dimensions $D$. \changes{@Sam: (1) We should only compare with the extrapolated values from CCM (see Table 1)}. (c) The variation of local magnetization, $m$, with $J_d$ obtained from iPEPS calculations for various bond dimensions $D$. \changes{@Sam: Please add the CCM results for $m$?} (d) The canting angle, $\alpha$ as a function of $J_d$. The definition of the canting angle is shown in the inset.
%    \changes{@Sam: Please add the classical values as well. See text for formula. For CCM, if the extrapolated values are not available, please take only plot LSUB8, which is their biggest system size.}} \label{fig-gs_mag}
%\end{figure*}

\section{Ground state phase diagram}\label{sec:result_gs}
We begin by discussing two special points of ~\eqref{eq:hmail}, namely $J_d = 0$ and $J_d = 1$. In the first case, the Hamiltonian reduces to the uniform antiferromagnetic Heisenberg model on the ruby lattice (also known as the bounce lattice). This limit has been studied in several works as part of the phase diagram of the MLL~\cite{Farnell2011,Beck2024,Gresista2023}, all of which report a magnetically ordered ground state. However, Ref.~\cite{PhysRevB.101.115114} finds a valence bond crystal ground state via iPEPS simulations, with a ground-state energy of $-0.545089$ per site at $D=12$. In our iPEPS calculations, we obtain a ground-state energy of $-0.559448$, which is significantly lower than the previous iPEPS estimate, at the same bond dimension [see Fig.~\ref{fig-gs_mag_1}(a)]. This result is in excellent agreement with the results obtained from CCM and other numerical methods. In Fig.~\ref{fig-gs_mag_1}(c), we show the variation of the average local magnetic moment,
\[
m = \frac{1}{6} \sum_{i=1}^{6} |\langle \vec{S}_i \rangle| ,
\]
where $i$ runs over the magnetic unit cell, as a function of $1/D$. The data indicates a finite $m$ as $D \to \infty$. 

The $J_d = 1$ limit has been studied using several numerical approaches~\cite{Farnell2011,maple_richter,Beck2024,Gresista2023,Hutsak_maple}, all of which report magnetic ordering. Our iPEPS calculations confirm this behavior. In Fig.~\ref{fig-gs_mag_1}(b), we show the ground-state energy as a function of $1/D$, compared with results from other numerical methods. In Fig.~\ref{fig-gs_mag_1}(d), we display the corresponding dependence of $m$ on $D$, indicating magnetic ordering in the $D \to \infty$ limit.

The spin moments obtained in our calculations for both $J_d = 0$ and $J_d = 1$ are small and comparable to the orderer magnetic moment found in the spin-$1/2$ nearest-neighbor Heisenberg model on the triangular lattice~\cite{Hasik_tri}.
%compared to those in other Archimedean lattices that exhibit magnetic ordering in the ground state. 
In the ideal maple-leaf limit, the magnetic moment is $m = 0.1697$ (at $D = 12$), while for the ruby lattice the moment is slightly larger, $m = 0.1856$. The local magnetizations and ground-state energies at both limits are summarized in Table~\ref{tab:energy}, together with a comparison to results obtained using other numerical techniques. The numbers corresponding to $m$ in Table~\ref{tab:energy} should be considered as upper bounds, since they decrease systematically with decreasing $1/D$. However, even under conservative linear extrapolations, which likely provide lower bounds, the magnetic moment remains non-negligible.

\begin{table}[]
\begin{tabular}{l||c||c}
\hline
$e_g$      & $J_d=0$ & $J_d=1$ \\
\hline   
\hline  
CCM~\cite{Farnell2011}            & $-0.5605$        &  $-0.5279$       \\
NQS~\cite{Beck2024}         &  $-0.549897$       &  $-0.523571$       \\
iDMRG~\cite{Beck2024}                         &  $-0.56376$       &  $-0.53100$      \\
Entropy method \cite{Hutsak_maple}                &  -       &  $-0.5304(2)$     \\
iPEPS~\cite{PhysRevB.101.115114} &  $-0.545089$  & - \\
%iPEPS~\cite{schmoll2024bathingseacandidatequantum}  &  $-0.5639(20)$       &  $-0.5304(25)$      \\
\textbf{iPEPS} &   $\mathbf{-0.560526}$    &    $\mathbf{-0.528052}$      \\
\hline 
\\
\hline 
$m$         & $J_d=0$ & $J_d=1$ \\
\hline   
\hline  
CCM~\cite{Farnell2011}             & $0.1657$        &  $0.1690$       \\
NQS~\cite{Beck2024}         &  $0.151673$       &  -       \\
\textbf{iPEPS ($D=12$)} &   $\mathbf{0.1856}$       &         $\mathbf{0.1697}$ 
\end{tabular}
\caption{Comparison of ground state energy per site ($e_g$) and average local magnetic moment ($m$) obtained from our iPEPS simulations with other numerical approaches. The iPEPS results obtained in the present work are highlighted in bold. The values of $m$ correspond to $D = 12$ and therefore provide upper bounds for the local magnetizations.}\label{tab:energy}
\end{table}

We now turn to the remaining parts of the phase diagram. It can be shown analytically that the ground state of the Heisenberg model for $J_d \ge 2$ is an exact product state of singlets on the dimer bonds~\cite{Ghosh2022}. This analysis, however, is based on a variational argument and does not preclude the system from realizing the same product singlet state for $J_d < 2$. Our iPEPS calculations show that this state ceases to be the ground state at $J_d \approx 1.45$. For $J_d \lesssim 1.44$, the system develops a magnetic order, which persists down to $J_d = 0$. In Fig.~\ref{fig-gs_mag_2}(a), we show the evolution of the ground-state energy per site as a function of $J_d$. To pinpoint the phase transition, we continue to evaluate the energy of the canted-$120^\circ$ state even beyond its stability region, allowing a direct comparison with the energy of the dimer singlet state in the vicinity of the transition. This allows us to identify a first-order transition between the two phases occurring within $1.44 < J_d < 1.45$ [see the inset of Fig.~\ref{fig-gs_mag_2}(a)] and to rule out the presence of any intermediate phase. 

We also compute the average local magnetic moment $m$ throughout the magnetically ordered phase using iPEPS simulations. In Fig.~\ref{fig-gs_mag_2}(b), we show the behavior of $m$ as a function of $J_d$ for different bond dimensions $D$. The magnetic moment exhibits only a weak dependence on $J_d$. In this phase, the system adopts the classical canted-$120^\circ$ order shown in Fig.~\ref{fig-gs_mag_2}(c)~\cite{Fennel2011,Ghosh2022,Beck2024,Gresista2023}, which can be understood as a local $120^\circ$ order on individual $J_t$ triangles, with a relative canting of the magnetic moments between neighboring triangles. This type of magnetic order is unique to the MLL. The canting is determined by the competition between the $J_h$ and $J_d$ bonds and is therefore expected to vary with $J_d$. One way to parametrize this canting is via the relative angle $\alpha$ between the spin moments across the $J_h$ bonds [see Fig.~\ref{fig-gs_mag_2} (c)]. In the classical $S \to \infty$ limit, $\alpha$ can obtained from a Luttinger-Tisza analysis~\cite{Luttinger1946,Ghosh2022} as
\be\label{eq:alpha}
\alpha_\text{cl} = \pi - \arccos\Bigg( \frac{4 - J_d}{2\sqrt{J_d^2 - 2 J_d + 4}} \Bigg).
\ee
Quantum fluctuations are expected to renormalize this angle. 
To determine $\alpha$, we compute the components of the local spin moments on the sites within a single unit cell and extract the relative angles between moments across the $J_h$ bonds.
In Fig.~\ref{fig-gs_mag_2}(c), we present the canting angle $\alpha$ as a function of $J_d$ obtained from our iPEPS calculations. 
 We find that $\alpha$ deviates significantly from the classical prediction over most of the magnetically ordered phase, except at $J_d = 0$, where it coincides with $\alpha_\text{cl}$. The canting angles obtained from iPEPS are in excellent agreement with CCM results~\cite{Farnell2011}.

%\begin{figure}[t]
%    \includegraphics[width=0.9\columnwidth]{figure6.pdf}
%    \caption{\changes{If we have more data for $J_d>1.4$ please plot them as well.}} %\label{fig-gs_mag}
%\end{figure}

%\begin{figure}[t]
%    \includegraphics[width=0.98\columnwidth]{figure6.pdf}
%    \caption{
%} \label{fig-mag_plateau}
%\end{figure}

%\changes{\section{Maple Leaf Antiferromagnet in a Magnetic Field}\label{sec:result_mag}
%Next, we move to the exploration of the magnetization behavior for the exact dimer phase. It is well understood that in systems exact product singlet ground states will give rise to low-lying magnetization plateaus with and without a classical analog~\cite{frustrationbook,Hard-Boson-SSM,Corboz2014}. Such states are either understood as crystals of localized triplet excitations or exotic multi-triplet bound states with large unit-cells driven by repulsive interactions. Such expectations are also applicable to MLL dimer singlet phase~\cite{Ghosh2023}. The exploration of the magnetization behavior at the intermediate field regime has been investigates using a effective hard-core boson picture is explored in Ref.~\cite{Ghosh2023} providing magnetization plateaus at magnetization $m/m_s=1/6, 2/9, 2/7$, and $1/3$, where $m_s$ is the saturation magnetization. In the present manuscript our main focus is on the large magnetic field regime.}

%\changes{We find that between $m/m_s=1/3$ and the fully-polarized state, we only find one additional plateau state at $m/m_s=2/3$.}

\section{Conclusions and Discussions}\label{sec:conclusions}
We have explored the quantum phase diagram of the nearest-neighbor antiferromagnetic Heisenberg model on the maple-leaf lattice (MLL) using the iPEPS approach combined with a CTMRG scheme suited for $C_3$-symmetric lattices. Our study focuses on the $J$-$J_d$ model on the MLL, for which one expects an exact dimer singlet product ground state to emerge for sufficiently large $J_d$. We find that the exact dimer state stabilizes for $J_d/J \gtrsim 1.45$. Below this ratio, the system exhibits a canted-$120^\circ$ magnetically ordered phase with a small magnetic moment. We identify a first-order phase transition between these two phases. Moreover, we could resolve the spin canting between neighboring trimer motifs within the magnetically ordered phase. Our results are in agreement with previously reported numerical studies. We also emphasize that our simulations can capture ground states with unit cells of up to eighteen sites. Within this framework, we find no evidence for an additional intermediate phase, as reported in Ref.~\cite{Gresista2023} or suggested in Ref.~\cite{Beck2024}. We would further like to note that the CCM developed by Richter and coworkers ~\cite{Bishop2000,richter2007frustrated} provides a highly reliable and internally consistent description of the MLL antiferromagnet, and more generally of quantum spin systems with magnetically ordered phases, yielding quantitatively accurate energies and order parameters that serve as a valuable benchmark for modern numerical approaches~\cite{SciPostPhysLectNotes86}.

There are two recent preprints~\cite{schmoll2024bathingseacandidatequantum,schafer2025thermodynamicsheisenbergantiferromagnetmapleleaf} that also study the same model and report ground states that differ from ours, and, by extension, from earlier studies. The former employs iPEPS simulations with an automatic-differentiation-based optimization scheme and finds that the canted-$120^\circ$ order stabilizes only in a narrow region adjacent to the exact dimer phase. The remainder of the phase diagram down to $J_d = 0$ is reported to be magnetically disordered. The authors further claim that the ground state at $J_d = 0$ is a gapless spin liquid, while the state at $J_d = 1$ is a gapped spin liquid. The latter work ~\cite{schafer2025thermodynamicsheisenbergantiferromagnetmapleleaf} uses numerical linked-cluster expansions to study the model at $J_d = 1$ and finds a short-range correlated paramagnetic ground state composed of resonating hexagonal motifs. The energies obtained in these studies differ only marginally from those found in our simulations. These results suggest that the model hosts several closely competing low-energy states, reflecting the subtle selection of its ground state. Exploring the system along alternative parametric trajectories~\cite{Ghosh2024-al,Ghosh_Hida_Model_of_Kagome,Gembe_mll} or under different perturbations could uncover the underlying complexity of the MLL phase diagram.

\textit{Acknowledgments:}
This work is dedicated to the memory of Johannes Richter. The work at EPFL was supported by the Swiss National Science Foundation under Grant No.~212082. Numerical computations were performed using the facilities of the Scientific IT and Application Support Center of EPFL (SCITAS).

The iPEPS data are openly available \cite{Ghosh2025_data}.

\bibliography{Refs.bib}

\end{document}